# Vortex lattice symmetry break in nanostructured Nb thin films

Oleksandr V Dobrovolskiy[1], Evgeniya Begun[1], Michael Huth[1], Valerij A Shklovskij[2,3], and Menachem I Tsindlekht[4]

[1]*Physikalisches Institut, Goethe-Universität, 60438 Frankfurt am Main, Germany*
[2]*Institute of Theoretical Physics, NSC-KIPT, 61108 Kharkiv, Ukraine*
[3]*Physical Department, Kharkiv National University, 61077 Kharkiv, Ukraine and*
[4]*The Racah Insitute of Physics, the Hebrew University of Jerusalem, 91904 Jerusalem, Israel*

An advanced mask-less nanofabrication technique, focused electron beam-induced deposition (FEBID), has been employed on epitaxial Nb thin films for their ferromagnetic decoration by an array of Co stripes. These substantially modify the non-patterned films' superconducting properties, providing a washboard-like pinning potential landscape for the vortex motion. At small magnetic fields $B \leq 0.1$ T, vortex lattice matching effects have been investigated by magneto-transport measurements. Peculiarities in the field dependencies of the films resistivity $\rho(B)$ have been observed in particular for the vortex motion perpendicular to the Co stripes. The deduced field values correspond to the vortex lattice parameter matching the pinning structure's period, whereas no fields matching the stripe width have been observed, as it was reported previously [D. Jaque *et al.* Appl. Phys. Lett. **81**, 2851 (2002)] for Nb films grown on periodically distributed submicrometric lines of Ni.

PACS numbers:

Superconductivity and vortex matter in thin films can be substantially modified through nanostructuring and the use of a washboard pinning potential landscape[1]. In particular, the manipulation and control of fluxons is needed for the development of superconducting nanodevices operating with Abrikosov vortices similar to electrons in nano- and microelectronics[2]. So far, such devices can predictably operate only within a single-vortex approximation[3], i. e. at low magnetic fields $B$, preferably less than 100 mT. Thus a pinning potential period length ensues in the nanometer range for a triangular Abrikosov lattice, $a \simeq \sqrt{\Phi_0/B}$ with $\Phi_0 = 2.07 \times 10^{-15}$ Tm$^2$ being the magnetic flux quantum. Thereby, in order to experimentally study the nonlinear single-vortex dynamics with emphasis on washboard pinning potential-induced peculiarities in the magneto-resistive response of superconducting devices, advanced nanofabrication tools must be applied.

Previously, lithographic techniques[4] have been widely employed to provide periodically arranged pinning sites whose distribution and form are determined by a particular mask used. In the present work, we have employed a mask-less direct nanofabrication technology, gas-assisted focused electron beam-induced deposition (FEBID)[5] of Co. FEBID allows one to create washboard pinning potential nanostructures in conformity with a pre-defined pattern in the nanometer range. Particular advantages of FEBID include high resolution[6] (down to 30 nm lateral and 1 nm vertical in the present case of Co deposition[7]) needed to provide the vortex lattice matching at substantially smaller magnetic fields (10-30 mT) than those resulting from the lithographical structuring, namely usually 0.1-1 T[8], and the possibility to fabricate structures with large aspect ratio (height-to-width). We show by magneto-transport measurements that vortex lattice matching effects induced by FEBID nanostructures on epitaxial Nb thin films differ markedly from those observed for conventionally patterned superconductors[9].

Fig. 1 illustrates the sample preparation stages. Epitaxial Nb(110) oriented thin films have been prepared by DC magnetron sputtering onto a-plane (11$\bar{2}$0) sapphire substrates. Details concerning the film growth and characterization are given in Ref.[10]. The films were pre-patterned by standard photolithography followed by Ar ion-beam etching in order to define the pattern for a rotating-current scheme consisting of eight contacts. This scheme allows one to apply the transport current $|\mathbf{j}| = (j_x^2 + j_y^2)^{1/2}$ at any arbitrary angle $\alpha$ with respect to the Co stripes. The experimentally deducible quantities are the magneto-resistivities $\rho_x$ and $\rho_y$, with the total resistivity being $\rho(B, \alpha) = (\rho_x^2 + \rho_y^2)^{1/2}$. Though a quantitative analysis of the resistive response measured in such a geometry is complicated due to the inhomogeneous current distribution caused by the cross-strip geometry[11], a similar scheme was used earlier[9] for a sys-tematic search of vortex-pinning induced peculiarities in $\rho(B, \alpha)$ and is chosen for reference purposes. After pre-patterning, an array of Co stripes has been deposited onto the film surface in the inner part of the bridge. FEBID has been performed in a high-resolution scanning electron microscope (SEM) and represents the process by which a metal-organic precursor gas, in this case $Co_2(CO)_8$, adsorbed on a film surface, is dissociated in the focus of the electron beam into a permanent deposit and volatile components[12]. The deposit's composition was determined by energy-dispersive X-ray spectroscopy (EDX). The deposit is metallic Co ($\approx 85\%$ at.) with inclusions of O($\approx 9\%$ at.) and C($\approx 6\%$ at.) as a residue from the precursor. Atomic force microscopy (AFM) investigations of the nanofabricated profiles confirm the high-periodicity of the pre-defined structure.

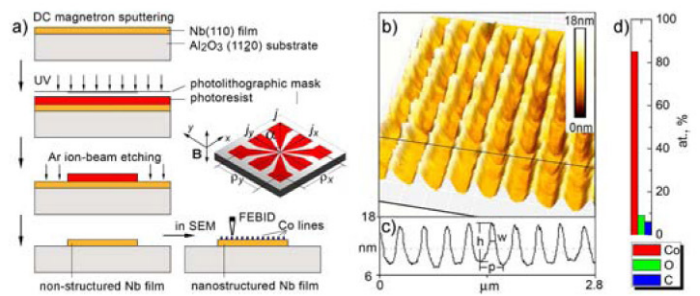

FIG. 1: (a) Preparation of FEBID-nanostructured Nb films. (b) AFM image taken in non-contact mode of the film surface after FEBID of Co. (c) Line scan of the AFM image, as indicated, with $h = 8$ nm, $w = 50$ nm, $p = 300$ nm denoting the structure's height, width and period, respectively. (d) Material composition in the fabricated nanostructures (electron beam parameters 20 kV/2.4 nA).

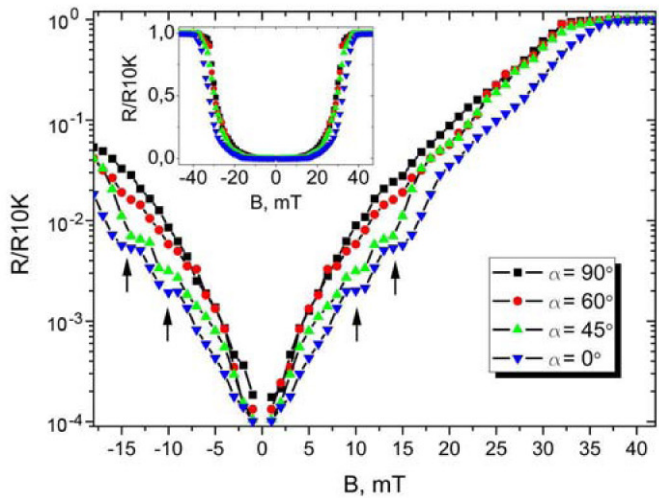

FIG. 2: Sample I (p=400 nm, w=50 nm). The field dependences of the normalized total resistivity $\rho(B)$ at $T = 0.99T_c$ as the transport current $j = 0.5$ kA/cm$^2$ is applied at different angles $\alpha$ with respect to the Co stripes, as indicated. Inset: the same plot for the whole field range in linear scale.

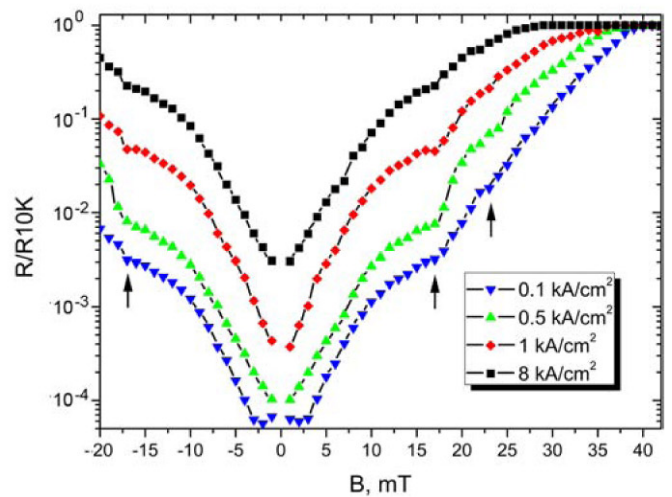

FIG. 3: Sample II (p=300 nm, w=50 nm). The field dependences of the normalized total resistivity $\rho(B)$ at $T = 0.99T_c$ for a set of the transport current values $j$, as indicated. The current is applied along the Co lines ($\alpha = 0°$).

Transport measurements have been performed in a cryostate equipped with a 9 T superconducting solenoid after careful compensation of the remanent field by using an additional magnet. During all the measurements the field **B** was applied perpendicular to the film plane in both, 'up' and 'down' directions. The temperature was kept as $0.99T_c$ with a stability better than 5 mK. The total current density $j$ was chosen as small as possible, i.e., $0.1, 0.5, 1, 8$ kA/cm$^2$, taking into account the vicinity to $T_c$. The 52nm and 54nm thick as-grown films (Sample I and II, respectively) demonstrated superconducting transitions in zero magnetic field at $T_c = 8.15$ K and 8.22 K, respectively. The transition width, using the $10\% - 90\%$ criterion, was found to be $< 0.1$ K. The residual resistivity ratios (RRR) were $R_{300K}/R_{9K} = 5$ and 6, respectively.

Fig. 2 shows the magnetic field dependence of the total magnetoresistivity $\rho(B, \alpha) = (\rho_x^2 + \rho_y^2)^{1/2}$ of Sam-ple I (p=400 nm, w=50 nm). The transport current was $j = 0.5$ kA/cm$^2$. The curves in Fig. 2 demonstrate several interesting features. First, with increasing $\alpha$, from $0°$ to $90°$, the superconducting transition shifts toward smaller fields. This is a clear signature of the guided vortex motion, i.e., the tendency of the vortices to move along the pinning potential channels rather than to overcome their potential barriers[13,14]. Here we assume a periodical suppression of the superconducting order parameter underneath the Co stripe as a result of the exchange coupling or proximity effect[15]. Second, the $\rho(B)$ curves exhibit peculiarities at 10 mT and 14 mT. If we assume a triangular vortex lattice, these fields correspond to the lattice parameters $a_L = \sqrt{\Phi_0/B} = 462$ nm and 400 nm, respectively. Whereas the lower field peculiarity at 10 mT ($a_L = 462$ nm) could be explained considering a triangular vortex lattice matching the structure period $a = a_L\sqrt{3}/2$, the peculiarity at 14 mT is not obvious and requires additional assumptions concerning the relative strengths of the vortex-vortex and vortex-structure interactions, as will be discussed below. Last, it should be noted that the peculiarities are most pronounced for $\alpha = 0°$ (vortex motion perpendicular to the Co stripes) and vanish for $\alpha = 90°$ (vortex motion parallel to the Co stripes). The qualitative behaviour of the $\rho(B)$ curves remains the same for other subcritical transport currents, namely $0.1, 1, 8$ kA/cm$^2$ (not shown). In order to check whether the stripe period or stripe width is chiefly responsible for the peculiarities observed, we prepared a second sample with the only difference in the period of the FEBID structure.

The dependence $\rho(B, j)$ of Sample II (p=300 nm, w=50 nm) is shown in Fig. 3. A pinning-induced anomaly in $\rho(B)$ is observed at 17 mT. The corresponding value of $a_L$ is 346 nm. An additional very weak anomaly in the $\rho(B)$ curves near 22-23 mT is detectable. Possible reasons will be discussed in the next paragraph. In addition, we have measured $\rho(B)$ at $T = 0.9T_c$ for both samples up to 1.5 T (not shown). No peculiarities in $\rho(B)$ corresponding to the $a_L$ value close to the stripe width have been observed. We assign this to the vortex lattice insensitivity to the Co stripes (as relatively weak pinning sites) of width shorter than the characteristic scale of the vortex-vortex interaction at high magnetic fields. One more reason could be a smoothed cross-section of the prepared nanoprofile. Having analyzed these findings, we now propose a qualitative scenario of the vortex lattice to penetrate the films while the magnetic field increases, as detailed next.

In Fig. 4 we plot $a_L(B) = \sqrt{\Phi_0/B}$ at magnetic fields between 0 and 22 mT. Possible vortex matching configurations (a-d) are shown for a 400 nm-periodic nanoprofile, as is representative for Sample I. The first lattice (sub-)matching (a,b) could cause peculiarities in the $\rho(B)$ curves at 2.4 mT and 3.2 mT, respectively. The experimentally measured curves $\rho(B)$ (see Fig. 2) reveal nothing special at these fields, however. We assign this to the small vortex density in the film at these fields. In accordance with charts (c,d) the next vortex lattice matching could occur at 10 mT and 13 mT for $a_L$=462 and 400 nm, respectively. Indeed, the experimentally observed peculiarities in Fig. 2 are in agreement with these values. As evident from Fig. 4.d for the case of a triangular vortex lattice, at the field of 13 mT ($a_L$=400 nm) additional unpinned vortices are in between the Co stripes. For this reason, this vortex lattice configuration is less stable than that at 10 mT (Fig. 4.c). If we assume a vortex lattice distortion from a triangular form, the presence of the peculiarities in $\rho(B)$ at 13 mT drawn for all $\alpha$ in Fig. 2 can be explained. If the vortex lattice was triangular at 13 mT, then there would be no peculiarities in $\rho(B)$. Instead, assuming a distorted triangular vortex lattice, which is known to occur[16] if the vortex-structure interaction is larger than the vortex-vortex interaction, the vortices feel relatively weak pinning force at $\alpha = 90°$ and are strongly pinned at $\alpha = 0°$. This is in agreement with the measured curves $\rho(B)$ in Fig. 2. The same arguments can explain the peculiarity suppression at 22 mT in the $\rho(B)$ curves for Sample II with the smaller period of 300 nm. Due to a shorter vortex-to-vortex distance, they interact more strongly and the pinning potential strength is too weak to provoke a vortex guiding along the Co stripes. The pinning potential efficiency is reduced as not all the vortices are pinned.

We want to stress, that the scenario described above gives only a qualitative account of the vortex lattice effects. For a quantitative description, temperature dependencies of the two main characteristic lengths which govern the superconductivity, namely the magnetic field penetration depth $\lambda(T)$ and coherence length $\xi(T)$, have to be taken into account. These are strongly dependent on the films' thickness and quality. Also, field dependencies of the critical current[17] at different $\alpha$ should be measured and analyzed. Calculations in the spirit of Martinoli's work[18] regarding the vortex lattice matching configurations and their relative stability for the nanostructures described here remain to be done and will be reported elsewhere.

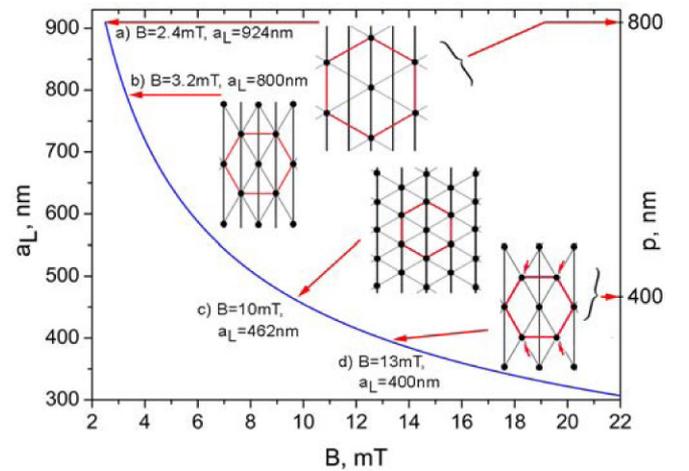

FIG. 4: Calculated value of the vortex lattice constant as a function of the magnetic field, $a_L = \sqrt{\Phi_0/B}$ (blue line). Possible vortex lattice matching configurations (cartoons a-b) are shown for the fields between 0 and 22 mT. Vertical black lines correspond to the Co stripes. Six nearest neighbors of a vortex are highlighted for leading the eyes. If the elastic energy of the vortex lattice becomes smaller than the pinning energy provided by the artificially created pinning sites, the vortex lattice gets distorted. Unstable vortices in between the Co stripes are shown by the arrows. Experimentally, peculiarities in $\rho(B)$ were observed only at $p = a$, where $a$ is the experimentally realized period of the washboard Co nanostructure (see text for details).

Concluding, linearly extended, periodic Co nanostripes prepared by FEBID allowed us to fabricate superconducting Nb thin films with washboard pinning potential landscapes for the vortices having a period which matches the vortex lattice parameter at small magnetic fields. This opens a pathway for a forthcoming experimental verification of the theoretical predictions for the guided vortex motion and the Hall-effect in superconductors with anisotropic pinning proposed by two of us[3].

The most important findings of the present work can be summarized as follows. (i) Peculiarities in $\rho(B)$ for FEBID-nanostructured Nb films were observed only when the vortex lattice parameter matches the nanostructure period. No matching effects corresponding to the Co stripe width have been observed. (ii) Peculiarities in $\rho(B)$ are more pronounced for the vortex motion perpendicular to the Co stripes, i.e., when the current flows parallel to them. (iii) A qualitative scenario for the vortex lattice to penetrate the film while the field increasing is considered. A distortion of the vortex lattice from a triangular form is proposed and experimentally supported by the presence of two matching fields observed in the $\rho(B)$ curves.

For a detailed understanding of the pinning mechanism provided by the Co-stripe decoration (proximity effect and/or exchange coupling) the micromagnetic state of the Co stripes needs to be investigated. Work along this line is currently under way.

The authors gratefully acknowledge financial support by the Beilstein-Insitut, Frankfurt/Main, Germany, within the research collaboration NanoBiC.

1. A. V. Silhanek, J. Van de Vondel, and V. V. Moshchalkov in *Nanoscience and Engineering in Superconductivity, Chap. 1, pp. 1-24* (Berlin Heidelberg: Springer-Verlag, 2010).

2. B. L. T. Plourde. IEEE Trans. Apl. Supercond. **19**, 3698 (2009).

3. V. A. Shklovskij and O. V. Dobrovolskiy, Phys.Rev. B **74**, 104511 (2006); **78**, 104526 (2008).

4. J. I. Martin, J. Nogues, K. Liu, J. L. Vicent, I. K. Schuller, J. Magn. Magn. Mater. **256**, 449 (2003).

5. I. Utke, P. Hoffmann, and J. Melngailis, J. Vac. Sci. Technol. B, **26**, 1197 (2008).

6. F. Porrati, R. Sachser, M. Strauss, I. Andrusenko, T. Gorelik, U. Kolb, L. Bayarjargal, B. Winkler and M. Huth Nanotechnology **21**, 375302 (2010); R. Sachser, F. Porrati, M. Huth, Phys. Rev. B **80**, 195416 (2009).

7. Depends on precursor. For selected precursors down to $\sim$ 10 nm at optimized beam parameters.

8. See, for instance, the work of Jaque et al.[9], who studied the field dependencies of resistivity on Nb films grown on periodically distributed submicrometric lines of Ni.

9. D. Jaque, E. M. Gonzalez, J. I. Martin, J. V. Anguita, J. L. Vicenta, Appl. Phys Lett. **81** 2851 (2002).

10. O. V. Dobrovolskiy, M. Huth and V. A. Shklovskij, J. Supercond. Nov. Magnet. DOI 10.1007/s10948-010-1055-7 (2010).

11. A. V. Silhanek, J. Van de Vondel, V. V. Moshchalkov, A. Leo, V. Metlushko, B. Ilic, V. R. Misko and F. M. Peeters, Appl. Phys. Lett. **92** 176101 (2008).

12. See e.g.: F. Porrati, R. Sachser, M. Huth, Nanotechn. **20**, 195301 (2009).

13. O. K. Soroka, M. Huth, and V. A. Shklovskij, Phys. Rev. B **76**, 014504 (2007).

14. O. V. Dobrovolskiy, M. Huth, and V. A. Shklovskij, Supercond. Sci. Technol. **23**, 125014 (2010).

15. A. I. Buzdin, Rev. Mod. Phys. **77**, 935 (2005).

16. P. Martinoli, J. L. Olsen, and J. R. Clem, J. Less-Com. Mater. **62**, 315 (1978).

17. N. Kokubo, R. Besseling, V. M. Vinokur, and P. H. Kes, Phys. Rev. Lett. **88**, 247004 (2002).

18. P. Martinoli, Phys. Ref. B. **17**, 1175 (1978).